\documentclass[a4paper]{article}

\usepackage{INTERSPEECH2019}
\usepackage[ruled]{algorithm2e}
\usepackage[utf8]{inputenc} % allow utf-8 input
\usepackage[T1]{fontenc}    % use 8-bit T1 fonts
\usepackage{hyperref}       % hyperlinks
\usepackage{url}            % simple URL typesetting
\usepackage{booktabs}       % professional-quality tables
\usepackage{amsfonts}       % blackboard math symbols
\usepackage{nicefrac}       % compact symbols for 1/2, etc.
\usepackage{microtype}      % microtypography

\usepackage{amsmath, amsfonts, amssymb, amsthm, bm}
\usepackage{graphicx}
\usepackage{framed}
\usepackage{hyperref}
\usepackage{bbm}
\usepackage{verbatim}
\usepackage{color}
\usepackage{multirow,tabularx}
\usepackage{soul}
\usepackage{subcaption}
\usepackage{caption}
\usepackage{commands}
\usepackage{array}
\usepackage{comment}

\title{Large-Scale Mixed-Bandwidth Deep Neural Network Acoustic Modeling for Automatic Speech Recognition}

\name{Khoi-Nguyen C. Mac$^{1}$, Xiaodong Cui$^{2}$, Wei Zhang$^{2}$ and Michael Picheny$^{2}$}
\address{$^{1}$ Department of Electrical and Computer Engineering, \\
         University of Illinois at Urbana-Champaign, IL 61801, USA \\
         $^{2}$ IBM Research AI \\
         IBM T. J. Watson Research Center, Yorktown Heights, NY 10598, USA }
\email{knmac@illinois.edu, \{cuix,weiz,picheny\}@us.ibm.com}

\begin{document}

\maketitle

\begin{abstract}
In automatic speech recognition (ASR), wideband (WB) and narrowband (NB) speech signals with different sampling rates typically use separate acoustic models. Therefore mixed-bandwidth (MB) acoustic modeling has important practical values for ASR system deployment.  In this paper, we extensively investigate large-scale MB deep neural network acoustic modeling for ASR using 1,150 hours of WB data and 2,300 hours of NB data. We study various MB strategies including downsampling, upsampling and bandwidth extension for MB acoustic modeling and evaluate their performance on 8 diverse WB and NB test sets from various application domains. To deal with the large amounts of training data, distributed training is carried out on multiple GPUs using synchronous data parallelism.
\end{abstract}

\section{Introduction}
\label{sec:intro}

Wideband (WB) and narrowband (NB) speech signals are two types of input signals that widely exist in speech-related applications. In automatic speech recognition (ASR), acoustic models are usually separately trained for WB and NB speech data given their distinct spectral characteristics under different sampling rates. From the system deployment's perspective, one acoustic model for both WB and NB speech would be greatly preferred. In this paper, we investigate mixed-bandwidth (MB) acoustic modeling using neural networks with deep architectures.

The goal of MB acoustic modeling is to converge the WB and NB speech to one bandwidth from which acoustic modeling is carried out. This could be accomplished either by downsampling or upsampling. In this paper we are interested in exploring MB acoustic modeling using deep neural networks (DNNs) and we are interested in seeking answers to the following questions: \textbf{(1) To converge to one bandwidth, which strategy is better, downsampling or upsampling? (2) how would direct mixing perform? (3) how would bandwidth extension (BWE) help in this case?} Furthermore, we are interested in real-world cases where large amounts of WB and NB training data are available and their amounts may be unbalanced. Specifically, we investigate MB deep convolutional neural network (CNN) acoustic models using 1,150 hours of WB speech and 2,300 hours of NB speech. We evaluate the ASR performance on a wide variety of WB and NB test sets collected from diverse scenarios.

To study the impact of BWE to MB acoustic modeling, we use a CNN with a VGG architecture \cite{Simonyan_VGG} to map the upsampled NB speech to WB speech. The CNN-based BWE network has its output connected to the input of an existing CNN acoustic model. It is discriminatively trained under the cross-entropy (CE) criterion.

Training deep CNN acoustic models using approximately 3,500 hours of speech data is computationally demanding. We resort to parallel computing for stochastic gradient descent (SGD) based network optimization with multiple GPUs. The system design and engineering consideration will be addressed in the system implementation.

The remainder of the paper is organized as follows. Section \ref{sec:related} will discuss the related work on MB acoustic modeling. Section \ref{sec:bwe} gives the mathematical formulation of BWE. Section \ref{sec:sys} is devoted to the system implementation including the model architectures and parallel computing. Experimental results are provided in Section \ref{sec:exp} followed by a discussion and summary in Section \ref{sec:dis}. 
\section{Related Work}
\label{sec:related}

MB acoustic modeling for ASR has been investigated previously under various conditions \cite{Seltzer_BWE,Li_BWE,Gao_BWE,Gao_MB}.

In \cite{Seltzer_BWE}, the NB data is used to leverage the training of WB acoustic models in a GMM-HMM framework and the missing components in upsampled NB speech features are dealt with using the expectation-maximization (EM) algorithm \cite{Dempster_EM} for the MB GMM-HMM acoustic models.  MB training in \cite{Li_BWE} follows a similar argument of \cite{Seltzer_BWE} as a missing feature problem but the problem is addressed in a DNN-HMM framework where no explicit BWE is assumed. Acoustic models in both \cite{Seltzer_BWE} and \cite{Li_BWE} are trained on a small amount training data (< 100 hours).

A joint MB training scheme is studied in \cite{Gao_BWE,Gao_MB} where BWE is used for MB. Specifically, a fully connected (FC) feedforward DNN is used to capture the BWE mapping from NB to WB speech with MMSE-based pretraining based on parallel speech data. Also, joint training is conducted to further improve the performance. There are 1,000 hours of WB data and 1,000 hours of NB data used in \cite{Gao_BWE} where experiments are conducted on one  WB and one NB test sets. There are 900 hours MB training data used in \cite{Gao_MB} (300 hours for 6KHz, 8KHz and 16KHz data respectively).

One difference among the previous literature is whether to assume an explicit BWE mapping and how it would help to the ultimate performance of MB models. Since DNNs are universal approximators, DNN-based BWE has been used for mapping between NB and WB speech. However, if we assume no explicit BWE network but only increase the capacity of MB DNNs, would it give rise to similar performance? We would like to design the experiments and evaluate with a competitive BWE network in this work.

\section{Bandwidth Extension}
\label{sec:bwe}

\begin{figure*}[t]
	\centering
	\includegraphics[width=13cm, height=4.25cm]{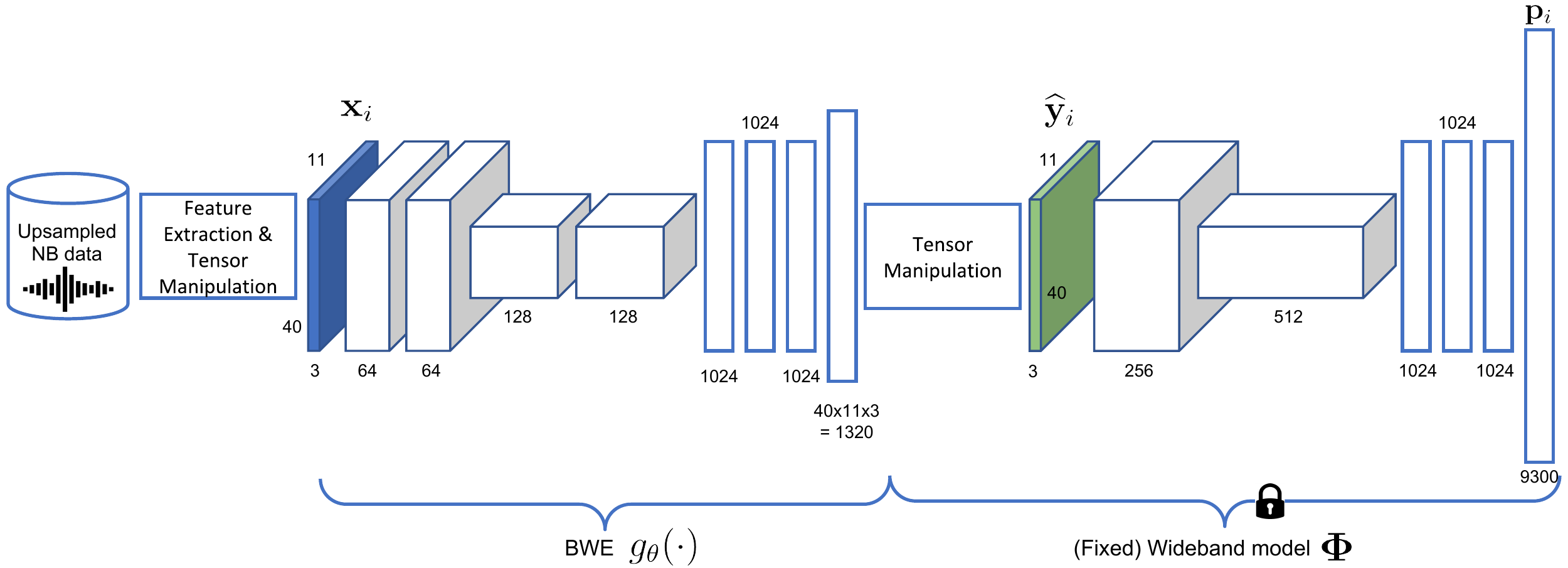}
	\caption{Illustration of the training of the BWE mapping. The mapping is realized as a CNN with a VGG architecture. Its output is connected to the WB CNN acoustic model after tensor manipulation. The WB CNN is fixed and the BWE CNN is optimized under the CE criterion.}
	\label{fig:discriminative}
\end{figure*}

BWE has been an active research topic in communication and acoustics processing. NB speech signals, such as telephony speech signals, suffer from degraded quality and intelligibility due to the lack of high frequency spectral information eliminated by the low-pass band limitation of communication channels. Over the years, extensive research has been carried out on BWE to compensate this degradation so as to improve the speech quality and intelligibility \cite{Iser_BWE,Prasad_BWE,Nagel_BWE,Pulakka_BWE,Liu_BWE,Ling_BWE}. BWE aims to estimate the missing high frequency spectral components and, therefore, effectively ``extend" the bandwidth of the speech signals. However, most of the work on BWE is optimized towards intelligibility, which may not be well aligned with ASR performance. In ASR-oriented BWE, we are interested in the mapping from a NB feature sequence to a WB feature sequence such as it improves the word error rate (WER).

Let $\mathbf{X} = \{\bm{x}_{1},\cdots,\bm{x}_{n}\}$ denote a sequence of $n$ NB features in certain feature domain, $\bm{x}_{i} \!\in\!\mathbb{R}^{d_{x}}$. We want to estimate a mapping function $f_{\theta}$ with some parameter $\theta$ to map the NB feature  sequence $\mathbf{X}$ to a WB feature sequence $\hat{\mathbf{Y}} = \{\hat{\bm{y}}_{1},\cdots,\hat{\bm{y}}_{n}\}$, $\hat{\bm{y}}_{i} \!\in\!\mathbb{R}^{d_{y}}$, where $\hat{\bm{y}}_{i} = f_{\theta}(\bm{x}_{i}).$
% \begin{align}
%      \hat{\bm{y}}_{i} = f_{\theta}(\bm{x}_{i}).
% \end{align}
A loss function
\begin{align}
    \mathcal{L}(\bm{l}_{i},\hat{\bm{y}}_{i}) \triangleq \mathcal{L}(\bm{l}_{i},f_{\theta}(\bm{x}_{i}))
\end{align}
is defined to measure the closeness of the mapped WB features $\hat{\bm{y}}_{i}$ and their labels $\bm{l}_{i}$. We want to optimize the parameter $\theta$ such that it minimizes the following empirical risk
\begin{align}
     \theta^{*} = \argmin_{\theta} \frac{1}{n}\sum_{i=1}^{n} \mathcal{L}(\bm{l}_{i},f_{\theta}(\bm{x}_{i})).
\end{align}
Depending on whether the problem is viewed as a regression or classification problem, the labels $\bm{l}_{i}$ are chosen differently.

BWE is typically treated as a regression problem and the mapping function is estimated under the MMSE criterion as follows \cite{Gao_BWE,Gao_MB,Likehuang_BWE}
\begin{align}
    \theta^{*} = \argmin_{\theta} \frac{1}{n}\sum_{i=1}^{n} \parallel\! \bm{y}_{i} - f_{\theta}(\bm{x}_{i}) \!\parallel_{2}^{2} \label{eqn:mse}
\end{align}
where the labels $\bm{l}_{i} = \bm{y_{i}}$, which is the ground truth WB counterpart of the NB speech features. This requires parallel WB and NB data and is usually accomplished by downsampling the WB speech to create the feature pairs. The mapping functions are learned in a reconstructive way to minimize the $L_{2}$ distance between the mapped NB features and their WB counterparts.

Since ASR is a classification problem by nature, it is desirable to have the BWE mapping estimated with a matched objective. In this paper, we choose another way of estimating the BWE mapping function. Suppose we have a WB neural network acoustic model $\mathbf{\Phi}$ which takes the WB speech features as input and outputs the posterior probabilities $p_{ik}$ with respect to context-dependent phone classes after the softmax layer for feature $i$ and class $k$. We use the BWE mapping function $g_{\theta}$ to map the upsampled NB speech features to WB features which are directly fed into the WB acoustic model $\mathbf{\Phi}$ to generate posteriors $\bm{p}_{i} = \mathbf{\Phi}(g_{\theta}(\bm{x}_{i}))$
where $\mathbf{\Phi}$ is fixed and $g_{\theta}$ is subject to optimization.  In this case, the mapping function is $f_{\theta} \triangleq \mathbf{\Phi} \circ g_{\theta}.$  The CE loss function is defined between the posterior probabilities $\bm{p}_{i}$ and the class labels $\bm{l}_{i}$:
\begin{align}
    \theta^{*} &= \argmin_{\theta} \frac{1}{n}\sum_{i} \mathcal{L}(\bm{l}_{i},\mathbf{\Phi}(g_{\theta}(\bm{x}_{i})) ) \\
    &= \argmin_{\theta} \frac{1}{n}\sum_{i,k} l_{ik}\log \frac{1}{p_{ik}}   \label{eqn:ce}
\end{align}
The labels $\bm{l}_{i}$ are generated by aligning the upsampled NB features against the WB acoustic model. The mapping function involves a composite of two CNNs. One is used to map the upsampled NB speech to WB which is subject to optimized and the other is an existing WB acoustic model which is fixed. This is illustrated in Fig \ref{fig:discriminative}.

In our pilot simulation experiments on 50-hours Broadcast News (BN50), we found that a DNN-based MMSE BWE did not help ASR performance when using a WB CNN model to decode the upsampled NB features mapped by it, which is worse than the NB baseline. Moreover, MMSE BWE relies on parallel data. Although this can be done by downsampling WB speech, it is an artificial setup. In real world such parallel data is difficult to collect. 
\section{System Implementation}
\label{sec:sys}

\subsection{Feature Space}

The WB speech is sampled at 16KHz while the NB speech at 8KHz. The input feature space consists of 40-dimensional logmel features after application of first a global cepstral mean normalization (CMN) and then an utterance-based CMN. There are three input feature maps to the CNNs, the static logmel features and their delta and double delta, all with a temporal context of 11 frames. For upsampled NB speech signals, they go through the WB Mel filter banks after upsampling in the time domain.

\subsection{Models}

\noindent\textbf{Acoustic models} \ \  CNN acoustic models are used for WB baseline, NB baseline and MB models, which have the same configuration. There are 2 convolutional layers and each convolutional layer is followed by a max-pooling layer. The first convolutional layer uses $5\!\times\!5$ kernels with a stride is $1\!\times\!1$ and padding $2\!\times\!2$. The second convolutional layer uses the same kernel, stride and padding sizes as those of the first convolutional layer. Both max-pooling layers use a kernel of $2\!\times\!2$ and stride of $2\times 2$. On top of the convolutional and pooling layers are 3 FC layers with 1,024 hidden units. All activation functions are Relu except the last FC layer which uses sigmoid. The output softmax layer has 9,300 output units. We investigate two model capacities in the experiments, one with 128 and 256 feature maps for the two respective convolutional layers and the other 256 and 512 feature maps.

\noindent\textbf{BWE Models} \ \  The BWE mapping network is also a CNN. The design of the network follows the VGG architecture that uses small convolutional kernels, small stride and small pooling kernels but, in the meantime, uses increased depth of the convolutional layers and  reduced max-pooling layers. Specifically, we use 4 convolutional layers, 2 max-pooling layers and 3 FC layers. Every 2 convolutional layers are followed by one max-pooling layer. The first 2 convolutional layers use $3\!\times\!3$ kernels with a stride $1\!\times\!1$ and padding $1\!\times\!1$. The second 2 convolutional layers again use $3\!\times\!3$ kernels with a stride $1\!\times\!1$ and padding $1\!\times\!1$. The 2 max-pooling layers use $2\!\times\!2$ kernels with a stride $1\!\times\!1$. The 3 FC layers have 1,024 hidden units. All activation functions are Relu except the last FC layer which uses tanh.  We also investigate two model capacities, one with 64 feature maps for the two convolutional layers and 128 feature maps for the next two convolutional layers, and the other 128 and 256 features maps. They are indicated when reporting the results.

\subsection{Distributed Training}
The networks are optimized under the CE criterion using the SGD algorithm. Learning rate starts as 0.01 for 10 epochs and is annealed by half every epoch for the next 10. We apply synchronous data parallelism on multi-GPUs (8 Nvidia v100s), within the same server, to accelerate training \cite{Zhang_GaDei}. To minimize parameters/gradients copy, we remap each trainable layer's gradient buffer to one consecutive region upfront. For each iteration, after each learner finishes backward propagation, we use NCCL\cite{NCCL} to sum all learners' gradients, via an Allreduce call, back to each learners' gradients region, before weights update. By doing this, 8-GPU training achieves almost 8x speedup. In our experiments, each GPU receives a mini-batch of size 512, bringing the total batch size per iteration to 4096. Another benefit of using multi-GPU is each learner effectively pre-load training data for others, which reduces I/O time to negligible.

% Mac's comment: these two paragraphs are redundant
% All models are trained on multiple GPUs in parallel. For each epoch, training data are divided into chunks, only one of which is loaded at a time. Each batch of the current data chunk is further split into $N$ sub-batches and sent to $N$ GPUs concurrently. One GPU is selected as the controller to manage checkpoints and training process. After back propagation, the computed gradients from all GPUs are combined together using NCCL all-reduce. This allows training process to be sped up almost linearly proportional to the number of GPU available.

%\begin{figure}
%	\centering
%	\includegraphics[width=\linewidth]{images/multigpus.pdf}
%	\caption{Multi-GPUs}
%	\label{fig:multigpus}
%\end{figure} 
\section{Experimental Results}
\label{sec:exp}

There are 1,150 hours of WB training data which consists of 420 hours of Broadcast News data, 450 hours of internal dictation data, 100 hours of meeting data, 140 hours of hospitality (travel and hotel reservation) data and 40 hours of accented data. There are 2,300 hours of NB training data which consists of 2,000 hours of Switchboard data and 300 hours of IBM call center data. In practice, we often find that the amounts of available WB and NB training data are unbalanced. In our case, the amounts of NB data is larger than that of the WB data.

We choose 4 WB test sets and 4 NB test sets for our experiments whose description and statistics are given in Table \ref{tab:datasets}. The decoding vocabulary comprises of 250K words and the language model (LM) is a 4-gram LM with modified Kneser-Ney Smoothing consisting of 200M n-grams. The LM training data is selected from a broad variety of sources. Stronger language modeling techniques such as RNN LMs will be investigated in the future. The word error rates (WERs) of various models on the 8 test sets are shown in Table \ref{tab:wer}.

\begin{table}[]
    \centering
    % \footnotesize
    \begin{tabular}{l|l|l|r}
            & & Description & Hours\\
        \hline \hline
        \multirow{4}{*}{WB} & WS1 & Dev04f test set from Broadcast News & 2.21  \\
                            & WS2 & Commercial services help desk       & 0.34  \\
                            & WS3 & Hospitality domain 1                & 1.21  \\
                            & WS4 & Hospitality domain 2                & 0.81  \\
        \hline
        \multirow{4}{*}{NB} & NS1 & Hub5-2000 test set from Switchboard & 2.10  \\
                            & NS2 & Technical support                   & 4.09  \\
                            & NS3 & Commercial services help desk       & 3.01  \\
                            & NS4 & Multi-domain command and control    & 12.78  \\
    \end{tabular}
    \caption{WB and NB datasets used for evaluation.}
    \label{tab:datasets} \vspace{-0.7cm}
\end{table}

\begin{table*}[htpb]
\centering
% \footnotesize
\begin{tabular}{l |c|c|c|c|>{\bfseries}c| c|c|c|c|>{\bfseries}c}
	& \multicolumn{5}{c|}{WB} & \multicolumn{5}{c}{NB} \\
	& WS1 & WS2 & WS3 & WS4 & Avg & NS1 & NS2 & NS3 & NS4 & Avg\\
	\hline
	\hline
	WB baseline  ([128,256])                & 15.4 & 14.9 &  9.1 & 29.2 & 17.2 & \underline{25.1} & \underline{39.0} & \underline{13.7} & \underline{22.0} & \underline{25.0} \\
	NB baseline ([128,256])                 & \underline{21.3} & \underline{16.8} & \underline{15.6} & \underline{40.5} & \underline{23.6} & 13.5 & 25.0 & 12.8 & 19.7 & 17.8 \\
	WB$\downarrow$  ([128,256])             & \underline{17.9} & \underline{17.5} & \underline{10.9} & \underline{33.9} & \underline{20.1} & 26.0 & 39.0 & 12.9 & 21.9 & 25.0 \\
	\hline
	DirectMix (WB+NB$\uparrow$,[128,256])    & 17.1 & 13.0 & 12.2 & 27.9 & 17.6 & 13.8 & 25.5 & 12.2 & 19.6 & 17.8 \\
	DirectMix (WB+NB$\uparrow$,[256,512])    & 16.5 & 12.8 & 11.8 & 28.8 & 17.5 & 13.4 & 25.2 & 11.8 & 19.2 & 17.4 \\
	DirectMix (WB$\downarrow$+NB,[128,256])  & 18.9 & 17.2 & 13.3 & 35.9 & 21.3 & 14.0 & 26.2 & 12.5 & 19.1 & 18.0 \\
	\hline
	BWE ([64,128])                          & -    & -    & -    & -    & -    & 15.2 & 27.8 & 12.4 & 20.2 & 18.9 \\
	BWE ([128,256])                         & -    & -    & -    & -    & -    & 14.9 & 27.4 & 12.2 & 20.0 & 18.6 \\
	nBWE ([64,128])                         & -    & -    & -    & -    & -    & 15.0 & 27.6 & 12.4 & 19.6 & 18.7 \\
	\hline
	Mix (WB+NB$\uparrow$+BWE, [128,256])    & 16.5 & 14.2 & 10.1 & 29.9 & 17.7 & 13.6 & 25.6 & 12.2 & 19.7 & 17.8 \\
	Mix (WB+NB$\uparrow$+BWE, [256,512])    & 16.0 & 14.6 &  9.7 & 29.9 & 17.6 & 13.7 & 25.4 & 12.2 & 19.6 & 17.7 \\
    Mix (WB+NB$\uparrow$+nBWE, [128,256])   & 16.4 & 14.3 & 10.0 & 30.9 & 17.9 & 13.7 & 25.6 & 12.1 & 19.5 & 17.7 \\
    \hline
    MixFT (WB+NB$\uparrow$+BWE, [128,256])  & 16.6 & 14.8 &  9.9 & 29.2 & 17.6 & 13.6 & 25.6 & 12.5 & 19.7 & 17.9 \\
    MixFT (WB+NB$\uparrow$+BWE, [256,512])  & 16.1 & 15.1 &  9.7 & 29.3 & 17.6 & 13.7 & 25.5 & 12.4 & 19.6 & 17.8 \\
    MixFT (WB+NB$\uparrow$+nBWE, [128,256]) & 16.2 & 14.4 &  9.8 & 30.3 & 17.7 & 13.6 & 25.4 & 12.0 & 19.6 & 17.7 \\
\end{tabular}
\caption{Word error rates (WERs) of WB, NB and MB models on 8 test sets. The WERs are reported in 5 blocks from top to bottom representing various experimental conditions.}
\label{tab:wer} \vspace{-0.5cm}
\end{table*}

\noindent\textbf{Baselines}  \ \ The WERs of WB and NB CNN baselines are shown in the first two rows in Table \ref{tab:wer}. The numbers of feature maps in the convolutional layers are indicated in the parentheses (e.g. [128,256] vs. [256, 512]). The underlined WERs indicate a change of sampling rate of the test data in order to be decoded by the acoustic model. For instance, the NB test sets are upsampled to decode against the WB CNN and vice versa. Obviously, without any compensation, mismatched test data and acoustic model give rise to significant degradation of the performance. (17.2\% $\rightarrow$ 23.6\% for WB test sets and 17.8\% $\rightarrow$ 25.0\% for NB test sets on average.)  We also carry out an experiment (third row) where only the WB training data is downsampled to train a CNN acoustic model which is used to decode the NB test sets. This model also gives a 25\% average WER which is significantly worse than the matched training with NB data only. The performance gap may be due to the mismatched data but also to the mismatched domains.

\noindent\textbf{Direct Mixing} \ \  The second block of Table \ref{tab:wer} presents the performance of MB models trained using the direct mixing strategy where WB data and upsampled NB data are mixed for the training of a CNN acoustic model.  The CNN model with [128,256] feature maps obtains about the same average WER as the NB CNN baseline (17.8\%) but slightly worse average WER than the WB CNN baseline (17.6\%). Since the amount of training data increases after mixing, it is reasonable to increase the capacity of the MB model. With doubled feature maps in the convolutional layers ([256,512]) the MB CNN has a lower average WER (17.4\%) on the NB test sets and only 0.3\% absolute worse on the WB test sets. On the other hand, however, if the MB model is trained using NB data and downsampled WB data, its performance is far inferior on both WB and NB test sets. Therefore, from the table we can tell that upsampling the NB data and then mixing with the WB data appears to be a better strategy for MB modeling.

\noindent\textbf{BWE} \ \ The third block of Table \ref{tab:wer} presents the performance of the proposed BWE approach. The VGG architecture of the BWE network is discriminatively trained with respect the WB CNN baseline model. With 64 feature maps in the first two convolutional layers and 128 feature maps in the second two convolutional layers, the BWE can significantly improve the average WER from 25.0\% to 18.9\%. If increase the network capacity with doubled feature maps, the average WER can be further improved to 18.6\%. The last row of this block shows the performance of BWE trained in a denoising manner (denoted nBWE) where zero-mean  Gaussian noise with variance of 0.01 is added to the input upsampled NB logmel features. It indicates that the denoising BWE can improve the generalization of the mapping and gives better performance under the same model capacity  (18.9\% $\rightarrow$ 18.7\%). Note that the BWE achieves improvement across all the four NB test sets against the WB CNN model compared to simple upsampling. In some test sets, BWE also yields better performance than the NB baseline. In the following experiments, we will stick to the BWE CNN configuration with the [64, 128] feature maps.

\noindent\textbf{Mixing with BWE} \ \ The fourth block of Table \ref{tab:wer} shows the performance of the mixing strategy of using WB data and BWE-mapped upsampled NB data from which the MB CNN models are trained. As shown by the table, the MB models further improve the WERs from the BWE alone. Using larger model capacity helps. Overall, it is slightly better than the direct mixing strategy when the model capacity is [128, 256] on the NB test sets. Denoising BWE helps the NB test sets but hurts the WB test sets.

\noindent\textbf{Fine-tune} \ \ The last block of Table \ref{tab:wer} shows the WERs after fine-tuning the mixing with BWE. In the fine-tuning, the output of the BWE CNN is connected to the input of the MB CNN which is fixed. The BWE CNN is fine-tuned with a smaller learning rate for 6 epochs. After that, another MB CNN is trained using the fine-tuned BWE with a smaller learning rate (1/10 of the original learning rate) for another 6 epochs. The improvement given by this finetuning, as can be observed from the table, is only marginal.

\section{Discussion and Summary}
\label{sec:dis}

As can be observed from the breakdown performance in Table \ref{tab:wer}, consistent improvements of one technique across all test sets are rarely observed. The conclusion drawn from one particular test set by one technique may not generalize to other test sets, although the average WERs can give us a good idea on the overall performance of certain technique. That is the reason we believe it is important to evaluate the BWE and mixed strategies extensively on diverse test sets from various domains and conditions.

In ASR applications, it is desirable to use one unified acoustic model for both WB and NB speech data. The experimental results in Sec.\ref{sec:exp} show that it is possible to train a MB model with an appropriate strategy. Upsampling the NB data appears to be more helpful than downsampling the WB data. In addition, direct mixing with appropriately increased model capacity, due to increased training data after mixing, can give competitive ASR performance compared to separate WB and NB models individually. In our investigated case, the best MB model yields lower average WER than the NB baseline and only slight degradation over the WB baseline. Although direct mixing assumes no explicit BWE, one would expect the DNNs will implicitly learn the mapping from the zero-padded upper frequency bins of NB speech to WB speech.

In our pilot experiments, MMSE-based BWE turned out not to be very helpful. Compared to the MMSE-based BWE, the proposed discriminatively trained BWE can significantly help when mapping NB speech to WB speech and decoded against WB models. Therefore, it is a competitive mapping function. However, when mixing WB speech and BWE-mapped NB speech for MB training, using BWE is not consistently better than direct mixing with increased model capacity. In addition, when considering BWE-based MB modeling, the BWE network should be treated as part of the MB model. Hence, its performance should be compared with direct mixing using networks with equivalent model capacity.

In summary, we have investigated in this paper the large-scale MB acoustic modeling with deep architectures for ASR. Extensive experiments were carried out to evaluate a variety of mixing strategies, including downsampling, upsampling and BWE, on diverse WB and NB test sets from various domains. Looking forward, with the success of deep generative models \cite{Goodfellow_GAN} and its applications to BWE \cite{Haws_BWECycleGan}, we hope to further improve BWE in the context of MB modeling for consistent superior performance.

\bibliographystyle{IEEEtran}

\bibliography{mixbwe}

\end{document}